\documentclass[
 reprint,
 nofootinbib,
 showkeys,
 amsmath,
 amssymb,
 aps,
 prb,
 floatfix,
 superscriptaddress,
 longbibliography
 ]{revtex4-2}

\usepackage{graphicx}
\usepackage{multirow}
\usepackage[dvipsnames]{xcolor}
\usepackage{dcolumn}
\usepackage{bm}
\usepackage{dsfont}
\usepackage{physics}
\usepackage{hyperref}

\usepackage{ulem}

\newcommand{\kk}{\mathbf{k}_{\perp}}

\begin{document}

\author{I.~V.~Iorsh}
\email{Contact author: ivan.iorsh@queensu.ca}
\affiliation{Department of Physics, Engineering Physics and Astronomy, Queen’s University, Kingston, Ontario K7L 3N6, Canada}

\title{Electron pairing by dispersive phonons in altermagnets: re-entrant superconductivity and continuous transition to finite momentum superconducting state.}

\date{\today}

\begin{abstract}
We consider an altermagnet subject to the electron attractive potential mediated by the bosonic excitations. While altermagnetism suppresses superconductivity, scattering of  electrons on the Fermi surface by thermal bosons suppresses altermagnetism. We show that this leads to the re-entrant superconductivity over temperature and to the stabilization of the Fulde–Ferrell–Larkin–Ovchinnikov (FFLO) finite momentum superconducting phase at low temperatures. The effect is mediated by the retardation effects and thus no $d$-wave pairing interaction is required.
\end{abstract}

\keywords{Superconductivity, altermagnet, Eliashberg theory}
\maketitle

\textit{Introduction.} Altermagnetism (AM) is a recently discovered class of magnetic ordering~\cite{AMGEN1,AMGEN2,AMGEN3,AMGEN4,AMGEN5}. The total magnetic moment in AMs vanishes  as in antiferromagnets, but magnetic sublattices are related by the rotation rather than by translation or inversion. The altermagnetism has been experimentally observed in various materials~\cite{AMEXP1,AMEXP2,AMEXP3,AMEXP4,AMEXP5,AMEXP6,AMEXP7,AMEXP8} including $\mathrm{RuO_2}$ and $\mathrm{MnTe}$.  the magnetic symmetries of altermagnets give rise of a plethora of novel and potentially useful effects in spin~\cite{Spintronics1,Spintronics2,Spintronics3,Spintronics4,Spintronics5,Spintronics6} and thermal~\cite{Therm2,Therm3} transport. 

Of particular interest is the interplay of altermagnetism and superconductivity: it has been anticipated that the altermagnetism may be used to realize Majorana zero modes~\cite{SC_app1,SC_Appl2}, novel types of Josephson junctions~\cite{AM_Josephson1,AM_josephson2} and superconducting diode effect~\cite{Diode_effect}. 

The altermagnetism may also lead to the emergence of finite momentum, FFLO~\cite{larkin1964nonuniform,fulde1964superconductivity} superconductivity~\cite{LOFF2} for the case of $d$-wave pairing mechanism. For the case of $s$-wave pairing in weak coupling regime it was shown in~\cite{Belzig} that altermagnetism suppresses superconducting transition temperature as in ferromagnets, but the transition remains of second order and thus no FFLO state is expected. At the same time, in~\cite{zhang2024finite} it was shown that if an altermagnet is proximitized with a $s$-wave superconductor, the induced cooper pairs gain finite momentum in the altermagnet.
\begin{figure}[!h]
\centering
\includegraphics[width=\columnwidth]{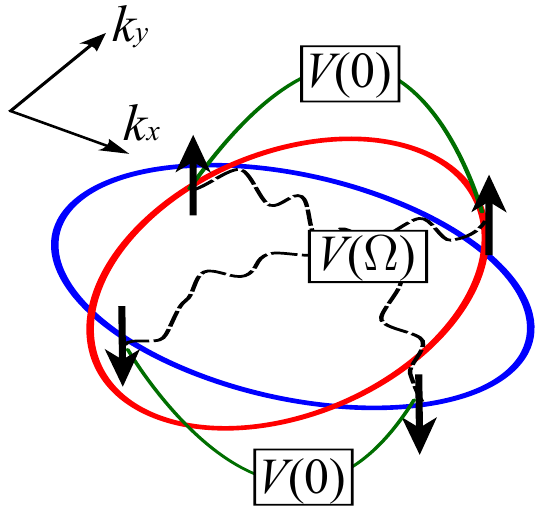}
\caption{\label{fig:Geom} Fermi contour of the two-dimensional altermagnet. Bosonic excitations with mediate the frequency dependent pairing potential $V(\Omega)$. Finite thermal concentration of the same bosonic excitations mediate the scattering of electrons on the Fermi surface   (labelled by $V(0)$) which  effectively suppresses the altermagnetic exchange field.}
\end{figure}

In this Letter we show that the account for the retardation effects in pairing drastically modifies the thermodynamics of altermagnetic superconductor. The essence of the effect can be visualized straightforwardly: bosonic excitations which facilitate the pairing have finite thermal concentration at finite temperature: these thermal bosons lead to the scattering of electrons over the Fermi surface which effectively suppresses the momentum dependent altermagnetic Zeeman field. We note that thermal bosons are irrelevant for the superconducting gap for the case of conventional s-wave superconductors~\cite{Eliashberg_1,Eliashberg_2}. They however become relevant in the altermagnetic case due to the spin polarization of the Fermi surface. The similar effect of scattering induced suppression of altermagnetism due to the static impurities has been recently proposed in~\cite{PhysRevB.111.L100502}, but the rate of scattering rate by thermal fluctuations is temperature dependent, which as we show leads to the destruction of conventional zero momentum supeconductivity at finite temperatures at the exchange fields slightly below the critical one, and the re-entrance to the superconducting regime at even lower temperatures with the stabilization of the finite momentum FFLO phase. 

The normal state  of the system is altermagnetic and is described by the Hamiltonian:
\begin{align}
H=\frac{p^2}{2m}+s_{AM}\sigma_z p_xp_y,
\end{align}
    Where $s_{AM}$ is the altermagnetic order parameter, and $\sigma_z$ is the Pauli matrix in the spin space. We also consider momentum independent pairing hamiltonian $V(i\omega_n-i\omega_{n'})$ where $\omega_{n}=\pi T(2n+1)$ are the Matsubara frequencies. We then apply the functional integral approach to derive the Eliashberg equations for the pairing vertex $\Phi(\omega_n)$ and electron self-energy $\tilde{\Sigma}(\omega_n)~$\cite{SuppLink1}
\begin{align}
&\Phi_n=T\pi\displaystyle\sum_{n'=-\infty}^{\infty} \tilde{V}_{n-n'}\frac{\Phi_{n'}\mathrm{K}(\kappa)}{(\pi/2)\sqrt{(|\Phi_{n'}|+\eta_0)^2+\tilde{\Sigma}_{n'}^2}}, \label{eq:Pn}\\
&\tilde{\Sigma}_n=\omega_n+\nonumber\\&{T\pi}\displaystyle\sum_{n'=-\infty}^{\infty} \tilde{V}_{n-n'}\frac{\mathrm{Im}\left[(\Phi+i\tilde{\Sigma}_{n'}-\eta_0)\Pi\left(\frac{2\eta_0}{\eta_0+\Phi_{n'}+i\tilde{\Sigma}_{n'}},\kappa\right)\right]}{(\pi/2)\sqrt{(|\Phi_{n'}|+\eta_0)^2+\tilde{\Sigma}_{n'}^2}}, \label{eq:Pn}
\end{align}
where
\begin{align}
\kappa=\frac{4\eta_0|\Phi_n|}{(|\Phi_{n}|+\eta_0)^2+\tilde{\Sigma}_n^2},
\end{align}
$\eta_0=s_{AM}p_{Fermi}^2$, $\mathrm{K}(x),{\Pi}(x)$ are complete elliptic integrals of the first and third kind, $\tilde{V}=\rho_0V$, where $\rho_0$ is the density of states on Fermi level, and $\tilde{\Sigma}_n=\omega_n +\frac{1}{2}[\Sigma^{\uparrow}(i\omega_n)-\Sigma^{\downarrow}(-i\omega_n)]$. The equation for the critical temperature is derived via the linearization of the self-consistent equation:
\begin{align}
&\frac{\Phi_n}{\pi T}=\displaystyle\sum_{n'} \tilde{V}_{nn'}\frac{\Phi_{n'}}{\sqrt{\eta_0^2+\tilde{\Sigma}_{n'}^2}},\frac{\tilde{\Sigma}_n}{\pi T}=\frac{\omega_n}{\pi T}+\displaystyle\sum_{n'}\tilde{V}_{nn'}\mathrm{sgn}(\omega_{n'}). \label{eq:linearized}
\end{align}

We consider the following form of the interaction potential  $\tilde{V}_{nn'}=V(i\omega_n-i\omega_{n'})$, where 
\begin{align}
V(i\Omega_m) = \int_0^{\infty}dy \frac{y\alpha^2F(y)}{y^2+\Omega_m^2},\quad \alpha^2F(y)=\lambda \gamma (y/\Lambda)^{\gamma}, y<\Lambda \label{eq:V}.
\end{align}
The Eliashberg spectral function $\alpha^2 F(y)=-\alpha^2F(-y)$ is defined by the characteristic energy scale $\Lambda$, dimensionless interaction strength $\lambda$ and exponent $\gamma$. Eq.~\eqref{eq:V} allows to describe a large class of models: for $\gamma\rightarrow \infty$ it corresponds to pairing via dispersionless Einstein phonons with frequency $\Lambda$, and $\lambda=(g/\Lambda)^2$ in this case. Moreover, for $\gamma\ge 1$, defining $\lambda=(g/\Lambda)^2$ and taking the limit of $\Lambda\rightarrow \infty$ we obtain the critical model $V(\Omega_m)=(g/\Omega)^2/(\gamma+1)$~\cite{PhysRevB.103.184508}. The dimensionless coupling $\lambda$ diverges in this case. The case of $\gamma=1/2$ has been reported to describe the pairing in dirty Fermi liquids~\cite{PhysRevB.110.155136}. Finally, the pairing by dispersive phonons is described by $\gamma=1$~\cite{kiessling2024boundstceliashbergtheory}. In this particular case, the integral in Eq.~\eqref{eq:V} can be taken analytically yielding:
\begin{align}
V(i\Omega_m)=\lambda\left[1- \arccot \left(\left|\frac{\Omega_m}{\Lambda}\right|\right)\left|\frac{\Omega_m}{\Lambda}\right|\right]
\end{align}
For $\Omega_m\ll \Lambda$ the asymptotics is $1-(\pi/2)|\Omega_m/\Lambda|$ and for $\Omega_m\gg \Lambda$ $V(i\Omega_m)=1/3(\Lambda/\Omega_m)^2$.
We further introduce the auxiliary function $Z_n=\tilde{\Sigma}_n/\omega_n$ and $\Delta_n=\Phi_n/Z_n$. The quasiparticle weight is given by $Z^{-1}_0$. We can rewrite the equation for $\Delta_n$
\begin{align}
&\Delta_n=T\pi \sum_{n'\neq n} {\tilde{V}_{nn'}}\left( \frac{\Delta_{n'}}{\sqrt{\omega_{n'}^2+\frac{\eta_0^2}{Z_{n'}^2}}}-\Delta_n\frac{\mathrm{sign}~\omega_{n'}}{\omega_{n}} \right).\label{eq:Delta_eq}
\end{align}
Thus the effective strength of altermagnetic order parameter on the pair function depends on the quasiparticle weight.
We note that in the absence of altermagnetism $\eta_0=0$, the Eq.~\eqref{eq:Delta_eq} decouples from the particle self energy $\tilde{\Sigma}$. Moreover, it can be seen that at $\eta_0$, the term $n=n'$ vanishes in right hand side of Eq.~\eqref{eq:Delta_eq}. This term corresponds to the contribution of scattering by thermal bosonic excitations on the Fermi surface, which is suppressed for the case of s-wave superconductors~\cite{Eliashberg_2}. Moreover, in the case of small self-energy corrections, $\Sigma(\omega_n)\ll \omega_n, \quad Z_n\approx 1$ the altermagnetic order plays the role of pair-breaking term which destroys superconductivity at some critical value of $\eta_0$: $T_c\rightarrow 0$ as $\eta_0\rightarrow \eta_{0}^c$~\cite{Belzig}. We can see in Eq.~\eqref{eq:Delta_eq} that in QCP limit when $g/\omega_0 \rightarrow \infty$, parameter $Z$ diverges leading to total suppression of the effect of altermagnetism. This is an analogue of the known effect of the robustness of superconductivity at QCP with respect to pair-breaking terms~\cite{PhysRevB.93.224514}.
\begin{figure}[!h]
\centering
\includegraphics[width=\columnwidth]{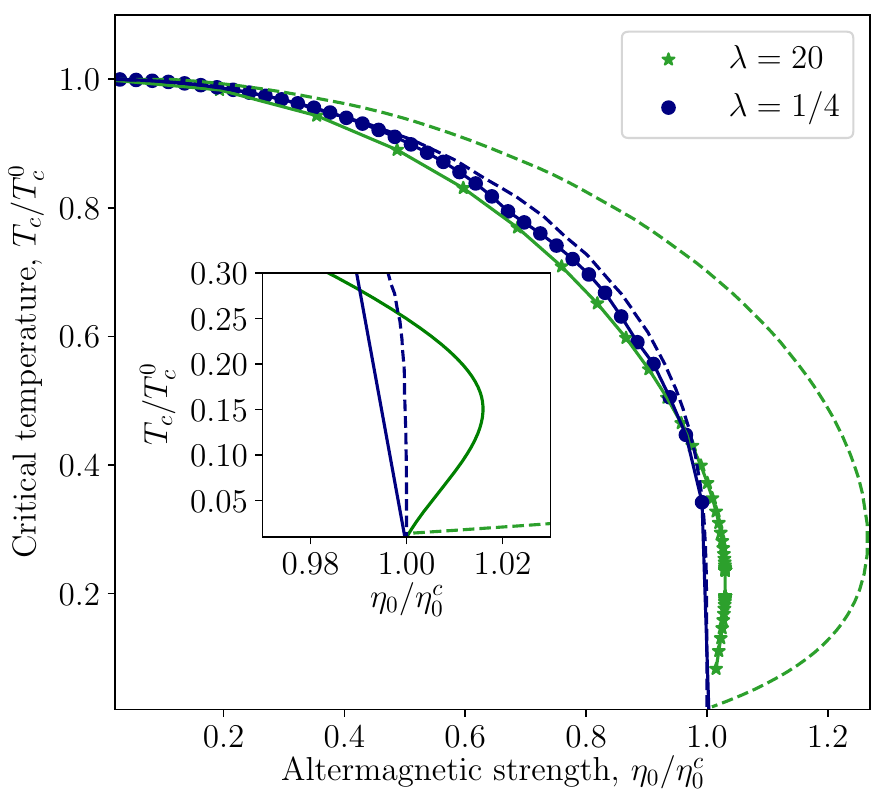}
\caption{\label{fig:TC} Dependence of the critical temperature $T_c$ normalized to that in the absence of altermagnetism $T_c^{0}$ on the altermagnetic order parameter $\eta_0$ normalized to $\eta_0^c$ at which $T_c=0$. Green lines correspond to $\lambda=20$ and navy to $\lambda=1/4$. Solid lines show the exact solution of the linearized self-consistent equation, while dashed are obtained using the upper bound for the largest eigenvalue in Eq.~\eqref{eq:Upper_Bound}. Inset shows the zoom in the vicinity of $\eta_0^{c}$}
\end{figure}

In Fig.~\ref{fig:TC} we plot the dependence of the critical temperature on the strength of the altermagnet order parameter $\eta_0$. The parameter $\gamma=1$. $T_c$ is normalized to $T_c^0$, the critical temperature at zero $\eta_0$ and $\eta_0$ is normalized to critical $\eta_0^c$ corresponding to vanishing $T_c$.  We see that for $\lambda=20$ corresponding to strong coupling regime, there is a region of temperatures in the vicinity of $T=0$, where the critical temperature increases as we increase $\eta_0$. In this region, the system is characterized by two critical temperatures: at $T=0$ the system is in the normal state; then at the lower critical temperature, the superconductivity is established. 
For comparison,  we also  plot the dependence for $\lambda=1/4$ corresponding to weak coupling: it can be seen that in this case the critical temperature always increases with decreasing $\eta_0$. The transition from monotonous to non-monotonous behavior occurs at $\lambda\approx 1$.

In order to derive anlaytical expression for critical temperature in the vicinity of the critical altermagnetic strength $\eta_0^c$ we note that the equation for the critical temperature, Eq.~\eqref{eq:linearized} can be approached as the problem of finding largest eigenvalue of corresponding linear operator $\mathcal{L}_{max}(T)$. Critical temperature is the solution of the equation $\mathcal{L}_{max}(T_c)=1$, and the system is in normal state when $\mathcal{L}_{max}<1$ and in superconducting when $\mathcal{L}_{max}>1$. Using the symmetry properties of the pairing vertex $\Phi_n=\Phi_{-(n+1)}$ we can rewite Eq.~\eqref{eq:linearized} as $\Phi_n=M_{nn'}\Phi_{n'}$ for $n,n'\ge 0$, where the infinite matrix $M_{nn'}$ is given by:
\begin{align}
M_{nn'} =\pi T\frac{\tilde{V}_{n-n'}+\tilde{V}_{n+n'+1}}{\sqrt{\tilde{\Sigma}^2_{n'}+\eta_0^2}}.
\end{align}
We can use a  upper bound for the largest eigenvalue~\cite{wolkowicz1980bounds}: 
\begin{align}
\mathcal{L}_{max}\le \mathcal{L}_0=\max_{n} \sum_{n'} M_{nn'}=\sum_{n'}M_{0n'} \label{eq:Upper_Bound}
\end{align}
The critical temperature dependence obtained with this upper bound is shown with dashed lines in Fig.~\ref{fig:TC}. As can be seen it reproduces the behaviour of the exact eigenvalue and demonstrates non-monotonic behaviour for $\lambda \gg 1$. We therefore can solve the equation $\mathcal{L}_0(T)=1$ for altermagnetic strength $\eta_0\approx \eta_0^{c}$, for which the critical temperature is exactly zero. We define $\eta_0=\eta_0^c+\delta\eta$ and solve the equation $\mathcal{L}_0 (T,\delta\eta)=1$ in the leading order in $T$ and $\delta\eta$. The zeroth order defines the critical $\eta_0^c$, $\eta_0^{c} \approx 2\Lambda e^{-(1+\lambda)/\lambda}$. To obtain the leading order expansion we use the Euler-Maclaurin~\cite{apostol1999elementary} formula for the approximation of sum by an integral
\begin{align}
\sum_{n'}M_{0n'}= M_{00} + \int_{1}^{\infty}M_{0n'} dn' + \frac{1}{2}M_{01}
\end{align}
In the leading order in $\delta\eta$ and $T$ we obtain:
\begin{align}
0=\zeta\frac{T_c^2}{(\eta_0^c)^2} - \frac{\delta\eta}{\eta_0^c}(1+\lambda)\left(\frac{\Lambda}{2\eta_0^c}+\frac{(\eta_0^c)^2}{6\Lambda^2}\right). \label{eq:Instab}
\end{align}
The parameter $\zeta$ can be approximated by:
\begin{align}
\zeta\approx \left(\frac{\Lambda}{\eta_0^c}\right)\left(6(\eta_0^c/\Lambda)^2-\lambda/(1+\lambda)\right).
\end{align}
The first, positive term in the bracket originates from the suppression of the effective exchange fiel by the scattering of thermal phonons, and the second, negative term, is usual for BCS superconductor, and corresponds to the suppression of superconductivity by thermal fluctuations. We note that if exponent $\gamma>1$ the first term would have $T^{\gamma+1}$ temperature dependence and thus will be subleading at low temperatures. For the weak coupling case, $\lambda\ll 1$, $\eta_0^c\approx T_c^0\approx \Lambda e^{-1/\lambda}$ and the equation can be approximated by $T_c/T_c^0\approx \sqrt{-\delta\eta/\Lambda}$, and thus critical exchange field decreases with temperature, as in BCS theory. In the strong coupling regime $\lambda \gg 1$, $\eta_0^c\approx T_c^0\approx \Lambda\sqrt{\lambda}$~\cite{kiessling2024boundstceliashbergtheory} and thus $T_c/T_c^0\approx \sqrt{\delta\eta/\eta_0^c}$. There is a critical coupling strength $\lambda_{cr}=O(1)$ where the parameter $\zeta$ changes sign.

The increase of the upper critical field in the strong coupling regime can be understood if we we recall that the thermal phonons play the role of the static impurities which facilitate scattering of the electrons on the Fermi surface. The characacteristic relaxation time due to theis process can be evaluted from the imaginary part of the self energy 
\begin{align}
\mathrm{Im}\Sigma(\omega,T)=-\pi \int_{-\infty}^{\infty} dy \alpha^2 F(y) (n_F(\omega+y)+n_B(y)),
\end{align}
where $n_F$ and $n_B$ are Fermi and Bose distributions, respectively. The inverse scattering rate $\tau^{-1}(T)=-\mathrm{Im}\Sigma$. For the case of vanishing real frequency $\omega=0$ and $T\ll \Lambda$ the integral simplifies to
\begin{align}
\tau^{-1}(T)=   \xi\Lambda\left(\frac{T}{\Lambda}\right)^{2}, 
\end{align}
where $\xi$ is the dimensionless constant of the order of unity. Thus, as temperature grows, the effective altermagnetic field is suppressed which increases critical field.  

The increase of the critical exchange field with temperature may be an indication of the instability towards finite momentum, Fulde–Ferrell–Larkin–Ovchinnikov superconducting state. We analyze the finite momentum linearized Eliashberg equations for this case. For the finite momentum pairing vertex $\Phi_{\mathbf{Q},n}$ the linearized Eliashberg equation yields~\cite{SuppLink2}:
\begin{align}
&{\Phi_\mathbf{Q},n}=\pi T\displaystyle\sum_{n'} \tilde{V}_{nn'} \times\nonumber\\&\int \frac{d\phi}{2\pi}\mathrm{Re}\frac{\Phi_{\mathbf{Q},n'}}{\sqrt{\left(\tilde{\Sigma}_{n'}+i\eta_0\cos(2\phi)+ivQ\cos(\phi-\varphi)\right)^2}}, \label{eq:FFLO}
\end{align}
Where $v$ is the Fermi velocity, and $\varphi$ is the angle between the crystal axis and the direction of the crystal field, $\mathbf{Q}=Q(\cos\varphi,\sin\varphi)$, and $\tilde{\Sigma}$ is the same as in Eqs.~\eqref{eq:linearized} In deriving Eq.~\eqref{eq:FFLO} we omitted all the terms proportional to $Q/k_F$, where $k_F$ is the Fermi wavevector to ensure the applicability of the Eliashberg theory~\cite{Eliashberg_2}. 

\begin{figure}[!h]
\centering
\includegraphics[width=\columnwidth]{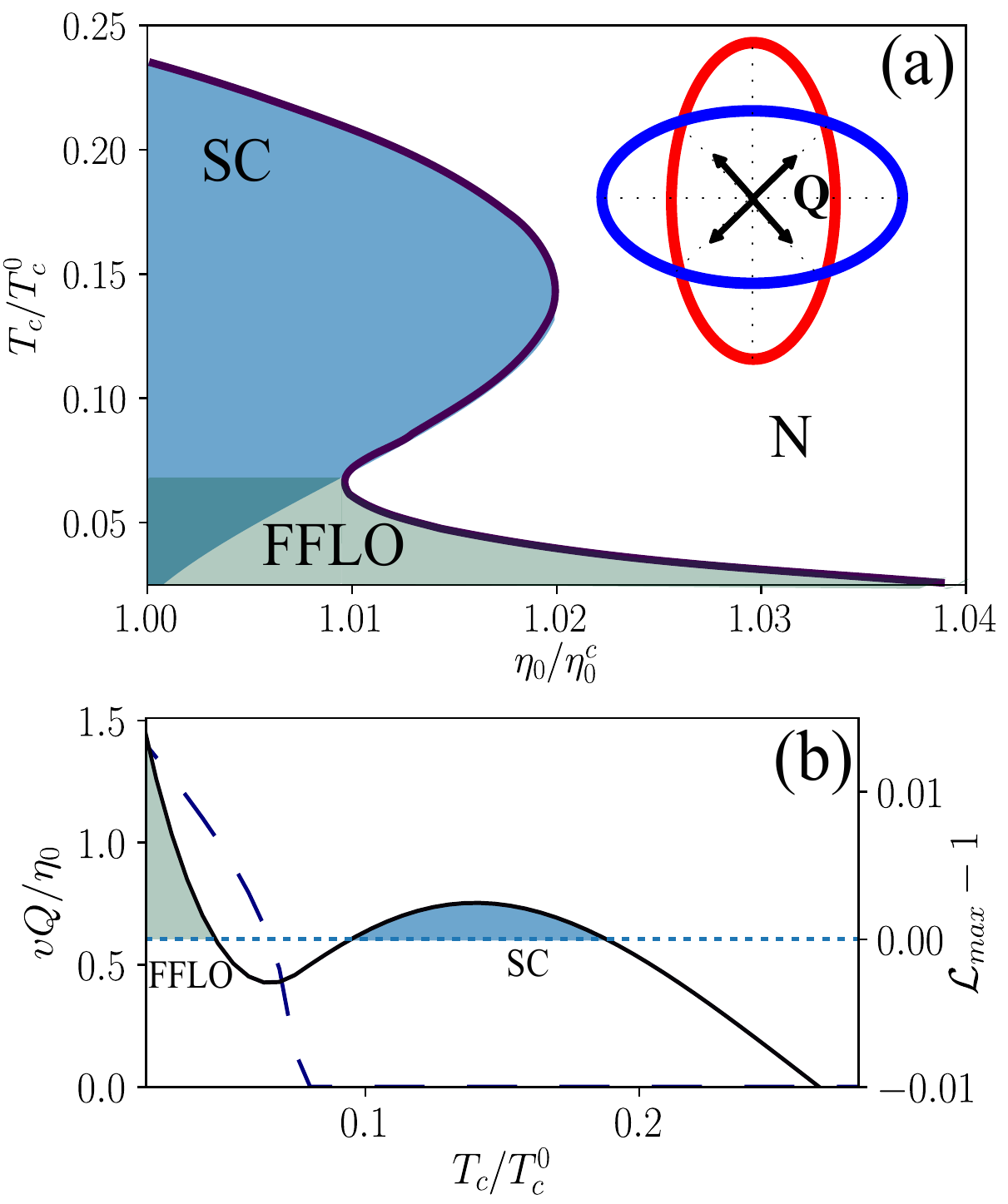}
\caption{\label{fig:3} (a) Phase diagram in the vicinity of the critical exchange field $\eta_0^c$ which includes normal, (N), superconducting at $Q=0$ (SC) and FFLO phase, superconducting at finite $\mathbf{Q}$. Inset shows the optimal directions of the $Q$ with respect to the crystal axes. (b) Cut of the phase diagram for $\eta_0=1.015~\eta_0^c$. Solid line corresponds to the value of the largest eigenvalue determining the onset of superconductivity (right axis) and navy dashed line shows the dependence of the optimal value of wavevector $Q$ (left axis). In the simulation $\lambda=20$.}
\end{figure}

Integration of Eq.~\eqref{eq:FFLO} results in an eigenvalue problem for the linear operator just as in Eqs.~\eqref{eq:linearized}, but now having $\mathbf{Q}$ as a parameter. We find the largest eigenvalue $\mathcal{L}_{max}(\mathbf{Q})$ and maximize it with respect to $\mathbf{Q}$. If this optimized $\mathcal{L}_{max}$ is larger than unity then the system is in superconducting state. The results of the calculation are shown in Fig.~\ref{fig:3}.

We find that at higher temperatures $\mathbf{Q}=0$ is the only stable configuration but as temperature decreases, a maximum at finite $Q$ emerges indicating the stabilization of the FFLO state. The direction of the optimal $\mathbf{Q}$ is always rotated by $\pi/4$ with respect to the crystal axis. There are thus 4 distinct directions of $\mathbf{Q}$. Within the linearized approach all four  directions are equivalent. When largest eiegenvalues for $Q=0$ and $Q\neq 0$, $\mathcal{L}_{max}(0)$ and $\mathcal{L}_{max}(\mathbf{Q})$ are both larger than unity, there is coexistence of phases in the system. In Fig.~\ref{fig:3}(b) a cut of phase diagram is shown for $\eta_0=1.015\eta_0^c$. We also plot the temperature dependence of the optmal $Q$ at the same plot. It can be seen that as temperature increases $Q$ reaches zero in the normal state region between FFLO and zero $Q$ superconductivity. In Fig.~\ref{fig:3}(a) it can be also seen that for low temperatures and $\eta_0/\eta_0^c<1.01$ the FFLO and $Q=0$ state coexist, and thus in this region transition to the FFLO state should be of the first order.

It can be seen that in Fig.~\ref{fig:3}(a) there still exists a region of instability where the critical exchange field increases with temperature. It is likely that in this region the system is unstable towards the other spatially inhomogeneous superconducting states with more complicated spatial dependence. We also note that the described behaviour does not require d-wave pairing discussed previously~\cite{LOFF2}. Thermodynamics of superconductor interfaced with altermagnet was considered within weak-coupling model in~\cite{Belzig} where no finite momentum instability was reported since as we show in Eq.~\eqref{eq:Instab} it requires strong coupling regime and pairing mediated by two-dimensional bosons with linear dispersion at small frequencies.

The proposed effect can be observed in a system comprising an altermagnet and a two-dimensional superconductor. The altermagnetic exchange field would be induced in the superconductor via the proximity effect. Among the possible candidates for superconductor, we could mention monolayer $\mathrm{Mg}_2\mathrm{B}_4\mathrm{C}_2$ where the strong coupling phonon-mediated superconductor was observed at 50 K~\cite{singh2022high}. At the same time, since the effect only requires linear dispersion of the pairing-mediating bosons, it could be observed in a number of materials belonging to rapidly expanding class of two-dimensional superconductors.

To conclude, we have shown that interplay of altermagnetism and elastic scattering of electrons by thermal bosons mediating pairing interaction may lead to non-trivial dependence of critical superconducting temperature on the atlermagnet exchange field and to instability towards finite momentum pairing. Importantly, the FFLO instability in altermagnet-superconductor structure can be realized with only $s$-wave pairing provided that the pairing falls to the strong coupling regime and the effect can be realized in the systems with conventional dispersive phonon mediated pairing.

The presented results highlight the importance of pairing retardation effects in the magnetic superconducintg structure and would find its application in spintronics and superconducting electronics.

\textbf{Acknowledgments}  I.I. acknowledges the
support of the Natural Sciences and Engineering Research
Council of Canada (NSERC) [Discovery Grant No. 2024-
05599]. I am also grateful to Prof. Erez Berg, Prof. Alexander Melnikov and Prof. Anton Samusev for helpful discussions.

\newpage
\onecolumngrid
\section*{Supplemental Material: Electron pairing by dispersive phonons in altermagnets: re-entrant superconductivity and continuous transition to finite momentum superconducting states}
\setcounter{equation}{0}
\renewcommand{\theequation}{S\arabic{equation}}

\section{Derivation of the Eliashberg equation for the altermagnetic superconductor}

We derive the Eliashberg equations from the path integral approach. We consider the general interaction $V(k-k')$ where $k$ is the four vector in Euclidian space $k=(i\omega_n,\mathbf{k})$. After integration over the fermion degrees of freedom we obtain the effective action:
\begin{align}
S(\bar{\Phi},\Phi,\Sigma) = -\mathrm{Tr}\ln ( \mathcal{G}^{-1})+\sum_{k,k'}V^{-1}(k-k')[\bar{\Phi}(k)\Phi(k')+\frac{1}{2}\Sigma^{\sigma}(k)\Sigma^{\sigma}(k')],
\end{align}
where $\sigma=(\uparrow,\downarrow)$. We assume only the singlet pairing, and we also assume that the interaction potential is real, i.e. $V(k-k')=V(k'-k)$. The Green's function in Namu space $\mathcal{G}$ is given by:
\begin{align}
\mathcal{G}^{-1}(k)= \begin{pmatrix} G^{-1}_{\uparrow}(k) & \Phi(k) & 0 & 0 \\ \bar{\Phi}(k) & - G^{-1}_{\downarrow}(-k) & 0 & 0 \\ 0& 0 & G^{-1}_{\downarrow}(k) & \Phi(-k) \\ 0 & 0 & \bar{\Phi}(-k) & -G^{-1}_{\uparrow}(-k) \end{pmatrix}, \label{eq:Gmatrix}
\end{align}
where $G^{-1}_{\uparrow\downarrow}(k)=i\omega_n-\xi_{\mathbf{k}}\mp \eta_{\mathbf{k}}-i\Sigma^{\uparrow\downarrow}(i\omega_n)$, and $\eta_{\mathbf{k}}$ is the altermangnet term $\eta=s_{AM}\cos(2\varphi)$.
Taking the variations with respect with respect to $\bar{\Phi}$ and $\Sigma$ we obtain the nonlinear equations for $\Phi$ and $\Sigma$:
\begin{align}
&\Phi(k) = \sum_{k'} V(k-k') \frac{\Phi(k')}{G^{-1}_{\uparrow}(k')G^{-1}_{\downarrow}(-k')+|\Phi(k')|^2}\\
&i\Sigma^{\uparrow}(k)= \sum_{k'} V(k-k') \frac{G^{-1}_{\downarrow}(-k')}{G^{-1}_{\uparrow}(k')G^{-1}_{\downarrow}(-k')+|\Phi(k')|^2},\\
&i\Sigma^{\downarrow}(-k)=\sum_{k'} V(k-k') \frac{G^{-1}_{\uparrow}(k')}{G^{-1}_{\uparrow}(k')G^{-1}_{\downarrow}(-k')+|\Phi(k')|^2}.
\end{align}
Now, if we do not have AM term, then we can assume $\Sigma^{\uparrow}=\Sigma^{\downarrow}$ and the equations in these case coincide with those presented in~\cite{PhysRevB.108.214508} and in \cite{PhysRevB.104.014513}.

In order to obtain more conventional form of the self-consistent equation we assume that both the interaction potential and the order parameters do not depend on the energy of electrons $\xi$ and there values are taken on the Fermi surface. Moreover, we separate the the component of the wavevector $\mathbf{k}_{\perp}$ parallel tangential to the Fermi surface.  We also introduce the modified order parameters
\begin{align}
&i\tilde{\Sigma}^{\sigma}(k)=i\omega_n-i\Sigma^{\sigma}(k)\\
& i\omega_n Z(k) = \frac{1}{2}i(\tilde{\Sigma}^{\uparrow}(k)-\tilde{\Sigma}^{\downarrow}(-k)),\\
& \chi(k) = \frac{1}{2}i(\tilde{\Sigma}^{\uparrow}(k)+\tilde{\Sigma}^{\downarrow}(-k)),\\
\end{align}
The equations then are rewritten as:
\begin{align}
&\Phi(k) = \sum_{k'} V(k-k') \frac{\Phi(k')}{(\chi(k')-\xi_{\mathbf{k'}})^2+(\omega_{n'}Z(k')+i\eta_{\mathbf{k'}})^2+|\Phi(k')|^2}\\
&Z(k)= 1+ \sum_{k'} V(k-k') \frac{Z(k')\omega_{n'}/\omega_n+i\eta/\omega_n}{(\chi(k')-\xi_{\mathbf{k'}})^2+(\omega_{n'}Z(k')+i\eta_{\mathbf{k'}})^2+|\Phi(k')|^2},\\
&\chi(k)=\sum_{k'} V(k-k') \frac{\xi-\chi(k')}{(\chi(k')-\xi_{\mathbf{k'}})^2+(\omega_{n'}Z(k')+i\eta_{\mathbf{k'}})^2+|\Phi(k')|^2}.
\end{align}
for the case of no AM, $\eta=0$ these equations coincide with those presented in\cite{marsiglio2020eliashberg}

If we now perform the integration over the energy $\xi$ from $-\infty$ to $\infty$ and assume particle hole symmetry, then $\chi$ is identically zero just as in\cite{marsiglio2020eliashberg}. The integration over energy yields 
\begin{align}
&\Phi(\mathbf{k}_{\perp},i\omega_n) = T\sum_{n'}\int \frac{d \mathbf{k}_{\perp'}}{(2\pi)^d\hbar v_F(\mathbf{k}_{\perp})} V(n-n',\kk-\kk') \frac{\Phi(n',\kk')}{\sqrt{(\omega_{n'}Z(n',\kk')+i\eta_{\kk'})^2+|\Phi(n',\kk')|^2}}\\
&Z(\mathbf{k}_{\perp},i\omega_n) = 1+T\sum_{n'}\int \frac{d \mathbf{k}_{\perp'}}{(2\pi)^d\hbar v_F(\mathbf{k}_{\perp})} V(n-n',\kk-\kk') \frac{Z(\kk',i\omega_{n'})\omega_{n'}/\omega_n+i\eta/\omega_n}{\sqrt{(\omega_{n'}Z(n',\kk')+i\eta_{\kk'})^2+|\Phi(n',\kk')|^2}}.
\end{align}
Written in this both $\Phi$ and $Z$ should be even functions of both $i\omega_n$ and $\mathbf{k}$. After taking the integral over $\xi$ we rewrite the equations for symmetric combinations $\Phi_c=\frac{1}{2}(\Phi(k)+\Phi(-k)$ to obtain
\begin{align}
&\Phi(\mathbf{k}_{\perp},i\omega_n) = \frac{T}{2}\sum_{n'}\int \frac{d \mathbf{k}_{\perp'}}{(2\pi)^d\hbar v_F(\mathbf{k}_{\perp})} V_{n-n',\kk-\kk'}\left[\frac{\Phi'}{\sqrt{(\omega_{n'}Z'+i\eta_{\kk'})^2+|\Phi'|^2}}+\eta\rightarrow-\eta\right],\\
&Z(\mathbf{k}_{\perp},i\omega_n) =1+ \frac{T}{2\omega_n}\sum_{n'}\int \frac{d \mathbf{k}_{\perp'}}{(2\pi)^d\hbar v_F(\mathbf{k}_{\perp})} V_{n-n',\kk-\kk'} \left[\frac{Z'\omega_{n'}+i\eta}{\sqrt{(\omega_{n'}Z'+i\eta_{\kk'})^2+|\Phi'|^2}}+\eta\rightarrow-\eta\right],
\end{align}
For  $Z=1$ and assume $V=V0$  the equation coincides with the one derived for the BCS altermagnetic superconductor~\cite{banerjee2024altermagnetic}.

We now consider  interaction potential $V$ accounting for retardation but momentum independent which can be written as:
\begin{align}
V=V(i\omega_n-i\omega_n');\quad 
\end{align}
 We then assume that the all order parameters are isotropic in the in the wavevector space. We also can take the integrals over $\phi$ analytically yielding:
\begin{align}
\frac{1}{2}\int \frac{d\varphi}{2\pi} \left[\frac{1}{\sqrt{(\omega_{n'}Z'+i\eta_0\cos2\varphi)^2+|\Phi'|^2}}+\eta\rightarrow-\eta\right]=\frac{2}{\pi} \frac{\mathrm{K}\left(\kappa\right)}{\sqrt{(|\Phi'|+\eta_0)^2+(\omega_{n'}Z')^2}},
\end{align}
where $\mathrm{K(x)}$ is the complete Elliptic integral, and 
\begin{align}
\kappa=\frac{4\eta_0|\Phi'|}{(|\Phi'|+\eta_0)^2+(\omega_{n'}Z')^2}
\end{align}
Another integral 
\begin{align}
\frac{1}{2}\int \frac{d\varphi}{2\pi} \left[\frac{i\eta_0\cos2\phi}{\sqrt{(\omega_{n'}Z'+i\eta_0\cos2\varphi)^2+|\Phi'|^2}}+\eta\rightarrow-\eta\right]=\frac{2}{\pi} \frac{\mathrm{Im}\left[(\eta_0-R)\Pi\left(\frac{2\eta_0}{\eta_0+R},\kappa\right)+R\mathrm{K}\left(\kappa\right)\right]}{\sqrt{(|\Phi'|+\eta_0)^2+(\omega_{n'}Z')^2}},
\end{align}
where $R=|\Phi|+i\omega_nZ_n$, and $\Pi(x,y)$ is the complete elliptic integral of the third kind.

The final set of equations to be solved ($g=\lambda_0 \rho$, where $\rho$ is the density of states):
\begin{align}
&\Phi_n=T\pi\displaystyle\sum_{n} V_{n-n'}\frac{\Phi_{n'}K(\kappa)2/\pi}{\sqrt{(|\Phi'|+\eta_0)^2+(\omega_{n'}Z')^2}},\\
&Z_n=1+\frac{T\pi}{\omega_n}\displaystyle\sum_{n'} V_{n-n'}\frac{\mathrm{Im}\left[(\Phi+i\omega_{n'}Z_{n'}-\eta_0)\Pi\left(\frac{2\eta_0}{\eta_0+\Phi_{n'}+i\omega_{n'}Z_{n'}},\kappa\right)\right]2/\pi}{\sqrt{(|\Phi'|+\eta_0)^2+(\omega_{n'}Z')^2}},
\end{align} 
Linearization of these equations results in the Eqs.~5 from the main text.

\section{Linearized Eliashberg  equations for FFLO state}
In this section we derive the linearized Eliashberg equations for the spatially inhomogeneous pairing vertex $\Phi_{\mathbf{Q}(k)}$. In our simulation we account only for the zero-th order term $\mathbf{Q}/k_F$ thus assuming that the modulation is very slow at the length scales of the inverse Fermi momentum. Within this approximation we neglect the vertex corrections and corrections due to non-circular Fermi surface. The effective action. Moreover, since we are intereetd in the linearized equations, we neglect the modification of the self energy $\Sigma$ by the pairing vertex and calculate via Eq. 5 in the main text. In this case the effective action can be written as:
\begin{align}
S(\bar{\Phi},\Phi) = -\mathrm{Tr}\ln ( \mathcal{G}_{0}^{-1}(1+\mathcal{G}_0\hat{\Phi}_{\mathbf{Q}}))+\sum_{k,k',\mathbf{Q}}V^{-1}(k-k')\bar{\Phi}_{\mathbf{Q}}(k)\Phi_{-\mathbf{Q}}(k'),
\end{align}
where $\mathcal{G}_0$ is the diagonal part of the matrix in Eq.~\eqref{eq:Gmatrix}, and matrix elements of the matrix $\hat{\Phi}_{\mathbf{Q}}$ are defined as $\hat{\Phi}_{\mathbf{Q},k_1,k_2}= \delta_{k_1,k2-\mathbf{Q}}\hat{\Phi}_0$, where $\hat{\Phi}_0$ is the off-diagonal part of the matrix in Eq.~\eqref{eq:Gmatrix}. The product in the logarithm can be decomposed in a sum of logarithm, where the first term $\ln \mathcal{G}_0^{-1}$ is a constant, and the second term can be expanded up to the second order in $\Phi$:
\begin{align}
-\mathrm{Tr}\ln (1+\mathcal{G}_0\hat{\Phi}_{\mathbf{Q}})\approx -\mathrm{Tr}[\mathcal{G}_0\hat{\Phi}_{\mathbf{Q}}]+\frac{1}{2}\mathrm{Tr}[\mathcal{G_0}\hat{\Phi}_{\mathbf{Q}}\mathcal{G}_0\hat{\Phi}_{\mathbf{Q}}].
\end{align}
The first term is zero, since $\mathcal{G}_0$ is diagonal and $\hat{\Phi}$ is off-diagonal. In calculation the trace in the second term we use the zeroth order approximation with respect to $Q/k_F$ yielding:
\begin{align}
&\Phi_{\mathbf{Q}}(k+\mathbf{Q})\approx \Phi_{\mathbf{Q}}(k),\\
&\eta_{\mathbf{k}+\mathbf{Q}}\approx \eta_{\mathbf{k}},\\
&\xi_{\mathbf{k+Q}}\approx \xi_{\mathbf{k}}+\mathbf{v\cdot Q},
\end{align}
where $\mathbf{v}$ is the vector of Fermi velocity.

Taking the trace and integrating over the energy $\xi$ yields

\begin{align}
\sum_{n,\mathbf{Q}} \int \frac{d\phi}{2\pi} \mathrm{Re}  \frac{\Phi_{n,\mathbf{Q}}\Phi_{n,-\mathbf{Q}}}{\sqrt{(\tilde{\Sigma}_n+i\eta_0\cos2\phi+vQ\cos(\phi+\varphi))^2}},
\end{align}
where $\varphi$ is the angle between $\mathbf{Q}$ and the crystal axis.
If we now take the variation of the action with respect to $\Phi_{n,\mathbf{Q}}$ and multimply by $V(k-k')$ we get Eq. 16 from the main text.

\end{document}